\documentclass[aps,pra,preprint,superscriptaddress,nofootinbib]{revtex4-2}

\usepackage{gensymb}
\usepackage{graphicx}
\usepackage{amsmath}
\usepackage{amssymb}
\usepackage{textcomp}
\usepackage{subcaption}
\usepackage{caption}
\usepackage{nicefrac}
\usepackage{upgreek}
\usepackage{wrapfig}
\usepackage{url}
\usepackage{setspace}

\usepackage{float}

\usepackage{tikz}
\usetikzlibrary{calc,arrows,decorations.pathmorphing,intersections}
\usepackage{pgfplots}
\pgfplotsset{width=10cm,compat=1.15}
\usepgfplotslibrary{fillbetween}
\usepackage{ctable}

\usepackage[T1]{fontenc}
\usepackage{lmodern}
\usepackage{listings}
\usepackage{framed}
\usepackage{tabularx}
\usepackage[justification=centering, width=0.8\textwidth]{caption}
\usepackage{subcaption}
\usepackage{color}

\makeatletter
\newcommand\BeraMonottfamily{%
  \def\fvm@Scale{0.85}
  \fontfamily{fvm}\selectfont
}
\makeatother

\newcommand{\ls}{1.0}

\lstnewenvironment{Jaqal}
{\lstset{
  language=C,
  showstringspaces=false,
  formfeed=\newpage,
  tabsize=2,
  commentstyle=\itshape,
  basicstyle=\linespread{\ls}\BeraMonottfamily, 
  breaklines=true,
  morekeywords={loop, macro, register, usepulse, from},
  xleftmargin=0.1\textwidth,
  xrightmargin=.25in,
  aboveskip=0pt,
  belowskip=0pt
}}
{}

\lstdefinestyle{jaqalstyle}{%
  language=C,
  showstringspaces=false,
  formfeed=\newpage,
  tabsize=2,
  commentstyle=\itshape,
  basicstyle=\linespread{\ls}\BeraMonottfamily, 
  breaklines=true,
  morekeywords={loop, macro, register, usepulse, from},
  xleftmargin=0.025\textwidth,
  xrightmargin=.05in
}

\lstdefinestyle{pythonblockstyle}{
  language=Python,
  showstringspaces=false,
  formfeed=\newpage,
  tabsize=4,
  commentstyle=\itshape,
  basicstyle=\linespread{\ls}\BeraMonottfamily, 
  emph={self},
  emphstyle=\itshape,
  breaklines=true,
  xleftmargin=0.025\textwidth,
  morekeywords={PulseData,models, lambda, forms},
  aboveskip=0pt,
  belowskip=0pt
}

\newcommand{\li}{\lstinline}

\newcommand{\pd}{\li{PulseData}~}

\newcommand{\code}[1]{
\lstinputlisting[aboveskip=0pt, belowskip=0pt]{#1}
}

\newcommand{\codeopts}[2]{
\lstinputlisting[#1]{#2}
}

\newcommand{\codespc}[1]{
\lstinputlisting{#1}
}

\newcommand{\jaqalcode}[1]{
\vspace{2em}
\begin{Jaqal}
#1
\end{Jaqal}
\vspace{2em}
}

\newcommand{\jaqalinput}[1]{
\lstinputlisting[style=jaqalstyle]{#1}
}

\newcommand{\pythoninput}[1]{
\lstinputlisting[style=pythonblockstyle]{#1}
}

\newcommand{\codeblock}[1]{
  \lstinputlisting[frame=single]{#1}
}

\newcommand{\parcodetab}[4]{
\vspace{.4\baselineskip}
\begin{tabular}{|l|l|}
\hline 
\parbox[c]{#3\textwidth}{
    \codespc{#1}
}
&
\parbox[c]{#4\textwidth}{
    \codespc{#2}
} \\
\hline
\end{tabular} 
}

\floatstyle{plain}
\newfloat{example}{thbp}{lop}
\floatname{example}{EX.}

\makeatletter
\patchcmd{\@footnotetext}
  {\setspace@singlespace}{0.8}
  {}{}
\makeatother

\newcolumntype{C}[1]{>{\centering\arraybackslash}p{#1}} 
\newcolumntype{R}[1]{>{\raggedleft\arraybackslash}p{#1}}                                                                  
\newcolumntype{L}[1]{>{\raggedright\arraybackslash}p{#1}}

\begin{document}


\title{JaqalPaw: A Guide to Defining Pulses and Waveforms for Jaqal}


\author{Daniel Lobser}
\email[]{dlobser@sandia.gov}
\homepage[]{https://qscout.sandia.gov}
\author{Joshua Goldberg}
\author{Andrew J. Landahl}
\affiliation{Sandia National Laboratories, Albuquerque, New Mexico 87123, USA}
\author{Peter Maunz}
\affiliation{Sandia National Laboratories, Albuquerque, New Mexico 87123, USA}
\affiliation{Currently at IonQ, College Park, Maryland 20740, USA}
\author{Benjamin C. A. Morrison}
\author{Kenneth Rudinger}
\author{Antonio Russo}
\author{Brandon Ruzic}
\author{Daniel Stick}
\author{Jay Van Der Wall}
\author{Susan M. Clark}
\affiliation{Sandia National Laboratories, Albuquerque, New Mexico 87123, USA}


\date{\today}

\begin{abstract}
One of the many challenges of developing an open user testbed such as QSCOUT~\cite{QSCOUTManual} is providing an interface that maintains simplicity without compromising expressibility or control.
This interface comprises two distinct elements: a quantum assembly language designed for specifying quantum circuits at the gate level, and a low-level counterpart used for describing gates in terms of waveforms that realize specific quantum operations.
Jaqal, or ``Just another quantum assembly language,'' is the language used in QSCOUT for gate-level descriptions of quantum circuits~\cite{Landahl2020,Morrison2020}.  JaqalPaw, or ``Jaqal pulses and waveforms,'' is its pulse-level counterpart.
This document concerns the latter, and presents a description of the tools needed for precisely defining the underlying waveforms associated with a gate primitive.
\end{abstract}

\maketitle
\newpage
\section{Background}

QSCOUT uses a linear chain of $^{171}\text{Yb}^+$ trapped ion qubits.
Coherent operations on these qubits involve manipulating both their internal spin states, as well as their collective motional states, via optical Raman transitions.
The specific experimental details of the hardware involved are described in the QSCOUT manual~\cite{QSCOUTManual}, but a few of these details are repeated here as they relate to certain aspects of the requirements for pulse-level control of these qubits.
Chief among these details is understanding the nature of these Raman transitions, and the pulsed laser used to drive them.

For the purposes of this manual, these elements can be distilled down to two basic rules of thumb:
\begin{itemize}
\item All operations must involve at least two tones to form a Raman transition.
\item The higher frequency tone in each Raman transition must employ a frequency feedback correction.
\end{itemize}
The specific details of how this frequency feedback system works, and why it is needed, are described in the QSCOUT manual~\cite[\S IV\,B, \S VI\,A\,2]{QSCOUTManual} and briefly overviewed in section~\ref{sec:freqfb}.
Beyond these basic rules, there are other subtleties in how these Raman beams can be configured, such as co- or counter-propagating for motionally insensitive or sensitive operations respectively.
Counter-propagating gates are achieved via the use of a global beam, which is driven by a specific hardware channel (channel zero by convention) and will be labeled \lstinline{GLOBAL_BEAM}. 
All other channels are used for individual qubit addressing.

From the perspective of engineering custom gates in terms of raw waveform data, it is imperative that the user have access to the latest calibration data.
Because calibration data is subject to change immediately prior to running a Jaqal program on the experimental apparatus, this information is exposed in the form of variables that are updated with the latest calibration data at run time.
Users are thus expected to reference these variables in their pulse definitions in order to ensure that their gates make use of the most recent calibrated data.

However, there are a number of details, independent of the specific knowledge of the calibration parameters, that are necessary for a basic understanding of how to use JaqalPaw to compose specific waveforms.
These details will be the primary focus of this manual, and will build from a kernel of basic elements that require no specific experimental knowledge of the QSCOUT system.

\section{Anatomy of a Gate Pulse Class}

Rather than developing a custom language for defining gates at the pulse level, JaqalPaw is written purely in Python and requires a simple set of conventions.
The first line in a Jaqal program should contain the desired set of gate definitions that need to be referenced.
This import has the form
\vspace*{1em}
\begin{lstlisting}
    from GateDefinitionFileName.GateDefinitionClassName usepulses *
\end{lstlisting}
This references a Python class which contains member functions that can {\it optionally} be exposed to Jaqal as long as the function name has a ``\lstinline{gate_}'' prefix.
These functions are then directly accessible in Jaqal and are identified by the function name where \lstinline{gate_} is implicitly stripped.
\begin{example}[H]
  \centering
  \parcodetab{code/anatomy/sidebyside.py.tex}{code/anatomy/sidebyside.jaqal.tex}{.4}{.5}
  \caption{\label{fig:jaqalpaw_jaqal}
  A side by side comparison of JaqalPaw (left) with Jaqal (right).
  The gate \li{gate_G} is accessed in Jaqal as \li{G}.
  }
\end{example}

The function's argument signature, with the exception of \lstinline{self}, can be used for passing input arguments from Jaqal.
In the above example, the gate \li{G} has a single input argument that corresponds to the target qubit.
The body of the gate definition function is up to the user, as JaqalPaw simply uses Python 3.
Likewise, any additional helper functions can be included and should simply leave off the \li{gate_} prefix.
Gate pulse definition classes do not use \li{__init__} methods, but instead make use of class-level attributes for passing in calibration data.
Calibration parameters always have a type annotation, and they will be overwritten at run time.
In order to access the standard calibration parameters, a user-defined class should inherit a standard QSCOUT gate class such as \li{qscout.v1.std}.
So an extension of the \li{GatePulse} class might look like the following

\begin{example}[H]
\begin{minipage}[t]{1.0\textwidth}
\codeopts{xleftmargin=0.1\textwidth,
          xrightmargin=.1\textwidth, 
          frame=single}{code/anatomy/simple_gp_class.py.tex}
\end{minipage}
\caption{\label{fig:basicGPClass}
A basic gate pulse class.
Gate definitions start with \li{gate_} and return a list of \pd objects.
Calibration parameters are class-level variables with type annotations.
}
\end{example}

This covers essentially all of the conventions used for constructing a simple gate pulse class.
The final detail is the return signature of a gate definition.
A gate's return signature must {\it always} be a list, and that list must only contain \li{PulseData} objects, even if there is only a single \li{PulseData} object in the list.

\section{The Pulse Data Class}

All of the low-level hardware control is exposed through the \li{PulseData} class, which is designed specifically to target the custom ``Octet'' coherent control hardware developed for QSCOUT.
This class is effectively a Python \li{dataclass}~\cite{Python3Dataclass}, in that it simply carries a set of parameters that are translated directly into a format sent to the coherent control hardware.
The full argument signature is shown in Fig.~\ref{code:pulsedataspec}, and the purpose and usage of each argument will be described in the remainder of this section.

\vspace*{2em}
\begin{figure}[H]
\centering
\codeblock{code/pulsedata/pulsedata.py.tex}
\caption{\label{code:pulsedataspec}
Full argument signature of \pd.
}
\end{figure}

\li{PulseData} at a bare minimum requires a channel and a duration, in which case the output pulse will effectively be a simple NOP for the given duration since the other parameters default to zero.
The remaining parameters can be separated into two categories: basic waveform data such as frequency, phase, and amplitude, and metadata used for controlling specific settings for the pulse.
Each \li{PulseData} object contains information for two independent tones that are digitally summed before being converted to an output waveform.
As opposed to an arbitrary waveform generator, these waveforms are always sinusoidal, but the frequency, phase, and amplitude of the oscillations are specified by the user.

\subsection{Controlling Frequency, Phase, and Amplitude}

The physical units for these inputs are listed in Table~\ref{tab:units}.
\begin{table}[h]
\begin{tabular}{|c|c|c|c|}
\hline
Parameter & Units & Allowed Range & Resolution \\
\hline
\hline
Time & s & $t \in [9.77~\text{ns}, 2684.35456~\text{s}]$ & 2.4414 ns\\
\hline
Frequency & Hz & $f \in [-409.6 \text{MHz}, 409.6 \text{MHz}]$ & 745.0581 $\mu\text{Hz}$\\
\hline
Phase & Degrees & $\theta \in [-\infty,\infty]$ & 3.2742e-10 deg.\\

\hline
Amplitude & Arb. & $\mathbb{R} \in [-100,100]$ & 6.1035e-3 \\
\hline
\end{tabular}
\caption{\label{tab:units}
The fundamental input units for frequency, phase, and amplitude in \li{PulseData}.
Note that the phase input is automatically converted modulo 360 such that $\theta \in [-180^\circ, 180^\circ)$.
Amplitude is specified for a single tone, however the sum of the amplitude for two tones on the same channel must obey this range.
}
\end{table}

These parameters can be controlled using three basic input types: a constant value, a list of values for discrete modulation, and a tuple of values for spline modulation.
For a constant value input, that value is applied for the specified duration of the pulse.
A very simple single-tone square pulse can be defined as shown in Ex.~\ref{ex:simplesqcode}.

\begin{example}[H]
  \centering
  \begin{minipage}{.8\textwidth}
    \code{code/pulsedata/gpplain.py.tex}
  \end{minipage}
\caption{\label{ex:simplesqcode}
A simple square pulse is defined in terms of constant parameters.
}
\end{example}

In Jaqal and JaqalPaw, gates are run back to back without any gaps unless strictly specified by the user.
This means that running the following Jaqal code will result in a single square pulse that has 4 times the duration of \li{G}.
\vspace{1em}
\begin{Jaqal}
loop 4 {
  G q[2]
}
\end{Jaqal}
\vspace{1em}
If a gap is desired, this could be implemented as either a separate \li{gap} gate with a short duration (e.g. one that returns \li{[PulseData(qubit, 0.25e-6)]}), or directly in the gate definition.
The back-to-back functionality applies to JaqalPaw in the same way it applies to Jaqal.
Namely, returning a list with multiple \pd objects on the same channel will be run in the order specified.
Thus, the following three cases are equivalent\footnote{To keep the code concise, only \li{amp0} is defined. \li{freq0} has been omitted, but would generally be needed to produce output on the hardware.}:

\begin{example}[H]
\centering
\vspace{.4\baselineskip}
\begin{tabular}{|l|l|}
\hline 
\parbox[c]{.6\textwidth}{\pythoninput{code/pulsedata/gpsquaregapsep.py.tex}} & \parbox[c]{.3\textwidth}{\jaqalinput{code/pulsedata/gpsquaregapsep.jaqal.tex} } \\
\hline
\parbox[c]{.6\textwidth}{\pythoninput{code/pulsedata/gpsquaregapsingle.py.tex}} & \parbox[c]{.3\textwidth}{\jaqalinput{code/pulsedata/gpsquaregapsingle.jaqal.tex} } \\
\hline
\parbox[c]{.6\textwidth}{\pythoninput{code/pulsedata/gpsquaregapmulti.py.tex}} & \parbox[c]{.3\textwidth}{\jaqalinput{code/pulsedata/gpsquaregapmulti.jaqal.tex} } \\
\hline
\end{tabular} 
\caption{\label{tab:differentloops}
Three equivalent ways of chaining pulses together. In these examples, gaps are explicitly added after each pulse.
}
\end{example}

When a series of discrete updates to a parameter is needed, the input to the parameter is a list of values.
Discrete modulations are treated as a series of back-to-back square pulses, where each pulse has the same time, $t'$, which evenly subdivides the total duration, $t$. 
In other words, a list with $N$ elements comprises $N$ pulses each with duration $t'=t/N$.

\begin{example}[H]
\vspace*{1em}
\captionsetup{width=.8\textwidth, justification=centering}

  \begin{minipage}[t]{.58\textwidth}
  \vspace{0pt}
  \code{code/pulsedata/gpdiscrete.py.tex}
  \end{minipage}%
  \begin{minipage}[t]{0.3\textwidth}
  \vspace{0pt}\raggedright

\begin{tikzpicture}
		\pgfplotsset{
			scale only axis,
			xmin=-0.5,xmax=4.5,
			ymin=-0.5,ymax=55.00000000000001,
            every axis plot/.append style={line join=round,line cap=round,clip=false}
		}

		\begin{axis}[
                        width=\linewidth,
			xlabel=Time ($\mu$s),
			ylabel=Amplitude (arb.),
			samples=100,
                        xticklabels={0.00,1.25,2.50,3.75,5.00},
                        xtick={0,1,2,3,4}
			]
\addplot[const plot, draw=black] coordinates {(0,10.0) (1,30.0)};
\addplot[const plot, draw=black] coordinates {(1,30.0) (2,20.0)};
\addplot[const plot, draw=black] coordinates {(2,20.0) (3,50.0)};
\addplot[const plot, draw=black] coordinates {(3,50.0) (4,50.0)};
	\end{axis}
\end{tikzpicture}
  \end{minipage}
  \hfill
\caption{\label{ex:discreteupdates}
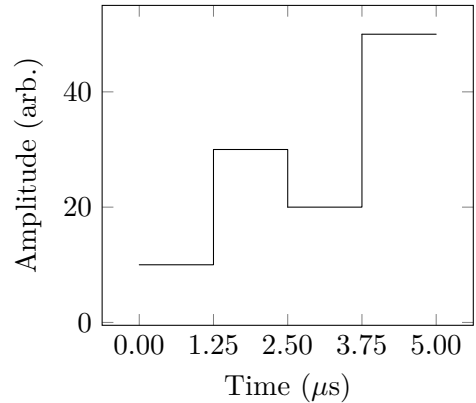
Discrete updates are represented as a list of inputs and are equally distributed across the duration of the pulse.
}
\end{example}

For continuous modulation via natural cubic splines, the input format simply changes from a list, \li{[...]}, to a tuple, \li{(...)}.
The input values are knots of the spline, and are also equally distributed over the duration of the pulse.
As opposed to the discrete case, a tuple of $k$ knots comprises $k-1$ sets of spline coefficients used for interpolation, resulting in a duration of $t'=t/(k-1)$ between knots. 

\begin{example}[H]
\vspace*{1em}

  \begin{minipage}[t]{.58\textwidth}
  \vspace{0pt}
  \code{code/pulsedata/gpsmooth.py.tex}
  \end{minipage}%
  \begin{minipage}[t]{0.3\textwidth}
  \vspace{0pt}\raggedright

\begin{tikzpicture}
		\pgfplotsset{
			scale only axis,
			xmin=-0.5,xmax=3.5,
			ymin=-0.5,ymax=55.00000000000001,
            every axis plot/.append style={line join=round,line cap=round,clip=false}
		}

		\begin{axis}[
                        width=\linewidth,
			xlabel=Time ($\mu$s),
			ylabel=Amplitude (arb.),
			samples=100,
                        xticklabels={0.00,1.67,3.33,5.00},
                        xtick={0,1,2,3}
			]
\addplot[][domain = 0:1]{-10.66667*(x-0)^3+-0.00000*(x-0)^2+30.66667*(x-0)^1+10.00000*(x-0)^0};
\addplot[][domain = 1:2]{23.33333*(x-1)^3+-32.00000*(x-1)^2+-1.33333*(x-1)^1+30.00000*(x-1)^0};
\addplot[][domain = 2:3]{-12.66667*(x-2)^3+38.00000*(x-2)^2+4.66667*(x-2)^1+20.00000*(x-2)^0};
	\end{axis}
\end{tikzpicture}
  \end{minipage}
  \hfill
\caption{\label{ex:smoothupdates}
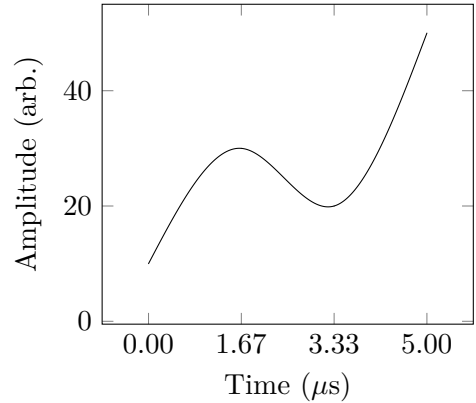
Smooth updates are represented as a tuple.
}
\end{example}

Multiple parameters can be modulated simultaneously and the length of the list/tuple can differ across all parameters. 
Asymmetry between the length of different modulation inputs is supported because the rules for equally distributing modulation data over the pulse time are handled separately for each parameter.

\begin{example}[H]
\vspace*{1em}

  \begin{minipage}[t]{.58\textwidth}
  \vspace{0pt}
  \code{code/pulsedata/gpSplineDiscreteComb.py.tex}
  \end{minipage}%
  \begin{minipage}[t]{0.3\textwidth}
  \vspace{0pt}\raggedright

    \begin{tikzpicture}
            \pgfplotsset{ 
                every axis legend/.append style={at={(0,1)},anchor=north west,font=\scriptsize},
                scale only axis,
                xmin=-0.5,xmax=7.5,
                ymin=-0.5,ymax=66.0,
                every axis plot/.append style={line join=round,line cap=round,clip=false}
            }

            \begin{axis}[
                width=\linewidth,
                xlabel=Time ($\mu$s),
                ylabel=Amplitude (arb.),
                samples=100,
                xticklabels={0.00,1.00,2.00,3.00,4.00,5.00},
                xtick={0.0,1.4,2.8,4.2,5.6,7.0}
                ]
    \addplot[draw=black][domain = 0.0:1.0]{-12.53865*(x-0.0)^3+0.00000*(x-0.0)^2+32.53865*(x-0.0)^1+10.00000*(x-0.0)^0};
\addplot[draw=black, forget plot][domain = 1.0:2.0]{32.69323*(x-1.0)^3+-37.61594*(x-1.0)^2+-5.07729*(x-1.0)^1+30.00000*(x-1.0)^0};
\addplot[draw=black, forget plot][domain = 2.0:3.0]{-48.23428*(x-2.0)^3+60.46376*(x-2.0)^2+17.77053*(x-2.0)^1+20.00000*(x-2.0)^0};
\addplot[draw=black, forget plot][domain = 3.0:4.0]{60.24390*(x-3.0)^3+-84.23909*(x-3.0)^2+-6.00481*(x-3.0)^1+50.00000*(x-3.0)^0};
\addplot[draw=black, forget plot][domain = 4.0:5.0]{-62.74133*(x-4.0)^3+96.49261*(x-4.0)^2+6.24871*(x-4.0)^1+20.00000*(x-4.0)^0};
\addplot[draw=black, forget plot][domain = 5.0:6.0]{50.72140*(x-5.0)^3+-91.73136*(x-5.0)^2+11.00996*(x-5.0)^1+60.00000*(x-5.0)^0};
\addplot[draw=black, forget plot][domain = 6.0:7.0]{-20.14428*(x-6.0)^3+60.43284*(x-6.0)^2+-20.28856*(x-6.0)^1+30.00000*(x-6.0)^0};
\addlegendentry{tone 0}
\addplot[const plot, draw=black,dashed] coordinates {(0.0,10.0) (1.4,30.0)};
\addplot[const plot, draw=black,dashed, forget plot] coordinates {(1.4,30.0) (2.8,20.0)};
\addplot[const plot, draw=black,dashed, forget plot] coordinates {(2.8,20.0) (4.2,30.0)};
\addplot[const plot, draw=black,dashed, forget plot] coordinates {(4.2,30.0) (5.6,50.0)};
\addplot[const plot, draw=black,dashed, forget plot] coordinates {(5.6,50.0) (7.0,50.0)};
\addlegendentry{tone 1}
        \end{axis}
    \end{tikzpicture}
    
  \end{minipage}
  \hfill
%
\caption{\label{ex:combinedtypes}
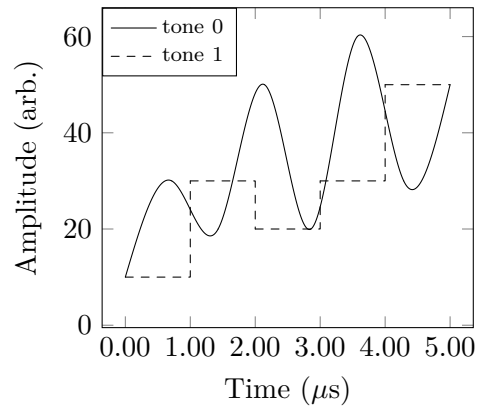
Different modulation types can be used across parameters.
}
\end{example}

Mixed modulation types on a single parameter are not currently supported in the form of a direct parameter input.
However, if one needs to combine modulation types on the same parameter, then multiple \li{PulseData} objects can be included in the return list to construct piecewise functions.
The following example illustrates how one can construct a piecewise function for a pulse, as well as how the natural cubic splines can be used to create linear ramps when only two knots are specified.

\begin{example}[H]
\vspace*{1em}

  \begin{minipage}[t]{.58\textwidth}
  \vspace{0pt}
  \code{code/pulsedata/gpPiecewise.py.tex}
  \end{minipage}%
  \begin{minipage}[t]{0.3\textwidth}
  \vspace{0pt}\raggedright

    \begin{tikzpicture}
            \pgfplotsset{ 
                scale only axis,
                xmin=-0.5,xmax=9.5,
                ymin=-0.5,ymax=55.00000000000001,
                every axis plot/.append style={line join=round,line cap=round,clip=false}
            }

            \begin{axis}[
                width=\linewidth,
                xlabel=Time ($\mu$s),
                ylabel=Amplitude (arb.),
                samples=100,
                xticklabels={0.00,2.00,4.00,6.00},
                xtick={0.0,3.0,6.0,9.0}
                ]
    \addplot[draw=black][domain = 0.0:1.0]{7.66667*(x-0.0)^3+0.00000*(x-0.0)^2+1.33333*(x-0.0)^1+0.00000*(x-0.0)^0};
\addplot[draw=black, forget plot][domain = 1.0:2.0]{-15.33333*(x-1.0)^3+23.00000*(x-1.0)^2+24.33333*(x-1.0)^1+9.00000*(x-1.0)^0};
\addplot[draw=black, forget plot][domain = 2.0:3.0]{7.66667*(x-2.0)^3+-23.00000*(x-2.0)^2+24.33333*(x-2.0)^1+41.00000*(x-2.0)^0};
\addplot[const plot, draw=black, forget plot] coordinates {(3.0,50.0) (4.0,50.0)};
\addplot[const plot, draw=black, forget plot] coordinates {(4.0,50.0) (5.0,50.0)};
\addplot[const plot, draw=black, forget plot] coordinates {(5.0,50.0) (6.0,50.0)};
\addplot[draw=black][domain = 6.0:7.0]{0.00000*(x-6.0)^3+0.00000*(x-6.0)^2+-16.66667*(x-6.0)^1+50.00000*(x-6.0)^0};
\addplot[draw=black, forget plot][domain = 7.0:8.0]{-0.00000*(x-7.0)^3+0.00000*(x-7.0)^2+-16.66667*(x-7.0)^1+33.33333*(x-7.0)^0};
\addplot[draw=black, forget plot][domain = 8.0:9.0]{0.00000*(x-8.0)^3+-0.00000*(x-8.0)^2+-16.66667*(x-8.0)^1+16.66667*(x-8.0)^0};
        \end{axis}
    \end{tikzpicture}
    
  \end{minipage}
  \hfill
\caption{\label{ex:piecewise}
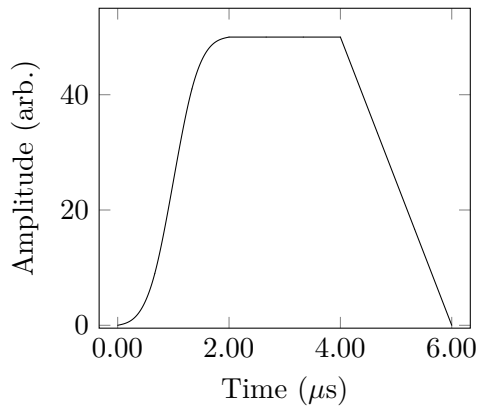
Piecewise functions can be constructed by chaining \pd objects together.
}
\end{example}

These parameters will be run in the order received on a per-channel basis.
If multiple channels are used, the data from each channel will be executed in parallel.
In situations where the total duration on each channel differs, channels will be padded with a NOP pulse at the end to ensure the duration is matched across all channels at run time.

\begin{example}[H]
\vspace*{1em}

  \begin{minipage}[t]{.58\textwidth}
  \vspace{0pt}
  \code{code/pulsedata/gpPiecewiseCounter.py.tex}
  \end{minipage}%
  \begin{minipage}[t]{0.3\textwidth}
  \vspace{0pt}\raggedright

    \begin{tikzpicture}[dbltransition/.style={black,<->,>=stealth',shorten >=0pt}]
            \pgfplotsset{ 
                every axis legend/.append style={at={(0,1)},anchor=north west,font=\scriptsize},
                scale only axis,
                xmin=-0.0,xmax=9.0,
                ymin=0,ymax=77.0,
                every axis plot/.append style={line join=round,line cap=round,clip=false}
            }

            \begin{axis}[
                clip=false,
                width=\linewidth,
                xlabel=Time ($\mu$s),
                ylabel=Amplitude (arb.),
                samples=100,
                xticklabels={0,1.50,3.00,4.50,6.00},
                xtick={0.0,2.25,4.5,6.75,9.0}
                ]
    \addplot[draw=black][domain = 0.0:1.0]{7.66667*((x-0.0)/(1.0))^3+0.00000*((x-0.0)/(1.0))^2+1.33333*((x-0.0)/(1.0))^1+0.00000*((x-0.0)/(1.0))^0};
\addplot[draw=black, forget plot][domain = 1.0:2.0]{-15.33333*((x-1.0)/(1.0))^3+23.00000*((x-1.0)/(1.0))^2+24.33333*((x-1.0)/(1.0))^1+9.00000*((x-1.0)/(1.0))^0};
\addplot[draw=black, forget plot][domain = 2.0:3.0]{7.66667*((x-2.0)/(1.0))^3+-23.00000*((x-2.0)/(1.0))^2+24.33333*((x-2.0)/(1.0))^1+41.00000*((x-2.0)/(1.0))^0};
\addplot[const plot, draw=black, forget plot] coordinates {(3.0,50.0) (4.0,50.0)};
\addplot[const plot, draw=black, forget plot] coordinates {(4.0,50.0) (5.0,50.0)};
\addplot[const plot, draw=black, forget plot] coordinates {(5.0,50.0) (6.0,50.0)};
\addplot[draw=black, forget plot][domain = 6.0:7.0]{0.00000*((x-6.0)/(1.0))^3+0.00000*((x-6.0)/(1.0))^2+-16.66667*((x-6.0)/(1.0))^1+50.00000*((x-6.0)/(1.0))^0};
\addplot[draw=black, forget plot][domain = 7.0:8.0]{-0.00000*((x-7.0)/(1.0))^3+0.00000*((x-7.0)/(1.0))^2+-16.66667*((x-7.0)/(1.0))^1+33.33333*((x-7.0)/(1.0))^0};
\addplot[draw=black, forget plot][domain = 8.0:9.0]{0.00000*((x-8.0)/(1.0))^3+-0.00000*((x-8.0)/(1.0))^2+-16.66667*((x-8.0)/(1.0))^1+16.66667*((x-8.0)/(1.0))^0};
\addlegendentry{indiv.}
\addplot[draw=black,dashed][domain = 0.0:2.25]{-14.66667*((x-0.0)/(2.25))^3+0.00000*((x-0.0)/(2.25))^2+44.66667*((x-0.0)/(2.25))^1+0.00000*((x-0.0)/(2.25))^0};
\addplot[draw=black,dashed, forget plot][domain = 2.25:4.5]{33.33333*((x-2.25)/(2.25))^3+-44.00000*((x-2.25)/(2.25))^2+0.66667*((x-2.25)/(2.25))^1+30.00000*((x-2.25)/(2.25))^0};
\addplot[draw=black,dashed, forget plot][domain = 4.5:6.75]{-18.66667*((x-4.5)/(2.25))^3+56.00000*((x-4.5)/(2.25))^2+12.66667*((x-4.5)/(2.25))^1+20.00000*((x-4.5)/(2.25))^0};
\addplot[const plot, draw=black,dashed, forget plot] coordinates {(9.0,0.0) (6.75,70.0)};
\addplot[const plot, draw=black,<-,>=stealth', forget plot] coordinates {(6.75,50.0) (9.0,50.0)};
\addplot[const plot, draw=black,<-,>=stealth', forget plot] coordinates {(9.,50.0) (6.75,50.0)}
        [pos segment=0,yshift=7pt,font=\footnotesize] node[pos=.5] {\textsc{nop}};
\addlegendentry{global}
        \end{axis}
    \end{tikzpicture}
    
  \end{minipage}
  \hfill
\caption{\label{ex:piecewisecounter}
Chaining \pd objects on different channels results in parallel execution.
Differences in cumulative duration will be padded with a NOP pulse.
}
\end{example}

\subsection{Virtual Rotations\label{sec:framerotations}}

In many systems, including QSCOUT, the ability to directly drive gates in all dimensions is often challenging, especially for individually addressed qubits.
Typically, two rotation axes are accessible, \textit{e.g.}, $\hat{x}$ and $\hat{y}$, in which case $\hat{z}$ rotations must be handled differently.

One approach is to treat $\hat{z}$ rotations virtually.
This generally involves redefining gates with adjusted phase offsets to account for the accumulated virtual phase.
However, the Octet hardware handles virtual rotations at the hardware level.
This allows gate definitions to be decoupled from the context dependency associated with virtual rotations.

The nomenclature used by JaqalPaw is ``frame rotation'', where the frame accumulates the input rotations until explicitly reset.
The input arguments to \pd are \li{framerot0} and \li{framerot1}. 
There is an important distinction between the \li{0} and \li{1} labels, which differ from the tone labels.

Frame rotations are intended to represent a virtual equivalent of a physical gate.
Because QSCOUT's gates use Raman transitions, the phase of the gate is determined by the {\it difference} between the phase of the two Raman tones.
Thus for co-propagating\footnote{Currently, QSCOUT is only using counter-propagating gates because of experimental requirements. However, co-propagating gates are mentioned here in terms of how one might implement them and some of the design choices that were made in terms of co-propagating gates.} gates, where both tones are set on the same channel, the frame rotation must only be applied to a single tone.
Two-qubit M\o lmer-S\o rensen gates often implement the motional sideband frequencies on a single channel, and the other leg of the Raman transition on the \li{GLOBAL_BEAM} channel.
The red and blue sideband tones independently form Raman transitions with the global beam.
This means that the frame rotation must be applied to both red and blue tones to achieve the proper virtual phase offset.

Other types of gate implementations may have an additional requirement that the virtual phase be inverted.
This condition may arise when the absorption and emission paths are swapped.
The effective phase of the gate relies on the phase difference between the two tones.
This means phase differences \li{phase0=0, phase1=90} and \li{phase0=-90, phase1=0} are essentially equivalent\footnote{The mean phase differs between the two cases, but the beat note at 12.642 GHz formed by the Raman beams will be equivalent between the two cases.\\}.

To account for all of these situations, the \li{framerot} parameters must be {\it forwarded} to the desired tones, and optionally inverted.
This forwarding and inversion can be handled on a per-tone basis, however the frame rotation information is independently tracked on a per-channel basis. 

Both the \li{framerot0} and \li{framerot1} inputs can be optionally forwarded to one or both tones.
Thus their treatment is slightly different than the other parameters such as \li{freq0} and \li{freq1}, which control tones 0 and 1 respectively.
\li{framerot1} can be used to track a second frame, such as one associated with a motional mode.
However, the hardware is currently limited to tracking two frames per channel, and doesn't impose any requirements about how they are implemented.
Rather, the convention used in QSCOUT is that the frame of the qubit spin is tracked in \li{frame0}.

Frame rotations obey the same input requirements as the \li{phase} inputs listed in table~\ref{tab:units}.
However, because the \li{framerot} inputs {\it accumulate} phase, this means that the inputs need to be treated a little bit differently.
Each application of a \pd object with a constant \li{framerot} will apply that phase with the beginning of the pulse by default.
Discrete modulations also obey the same behavior, consider the following definitions of \li{gate_G} which have the same behavior.

\begin{example}[H]
\vspace*{1em}

  \begin{minipage}[t]{.58\textwidth}
  \vspace{0pt}
  \code{code/pulsedata/gpFrameRotSingle.py.tex}
  \end{minipage}%
  \begin{minipage}[t]{0.3\textwidth}
  \vspace{0pt}\raggedright

    \begin{tikzpicture}
            \pgfplotsset{ 
                scale only axis,
                xmin=-0.5,xmax=3.5,
                ymin=-0.5,ymax=33.0,
                every axis plot/.append style={line join=round,line cap=round,clip=false}
            }

            \begin{axis}[
                width=\linewidth,
                xlabel=Time ($\mu$s),
                ylabel=Frame 0 (deg.),
                samples=100,
                xticklabels={0.00,1.00,2.00,3.00},
                xtick={0.0,1.0,2.0,3.0}
                ]
    
    \addplot[const plot, draw=black,dashed, forget plot] coordinates {(-1.0, 0.0) (0.0,0.0)};
    \addplot[const plot, draw=black] coordinates {(0.0,0.0) (0.0,10.0) (1.0,20.0)};
    \addplot[const plot, draw=black, forget plot] coordinates {(1.0,20.0) (2.0,30.0)};
    \addplot[const plot, draw=black, forget plot] coordinates {(2.0,30.0) (3.0,30.0)};
    \addplot[const plot, draw=black,dashed, forget plot] coordinates {(3.0,30.0) (4.0,30.0)};
        \end{axis}
    \end{tikzpicture}
   
  \end{minipage}
  \hfill
\caption{\label{ex:framerotsingle}
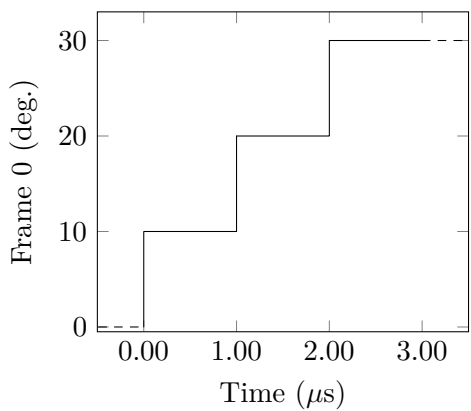
Frame rotation inputs are equivalent to phase, but their values accumulate.
}
\end{example}

Example~\ref{ex:framerotsingle} only shows the value of the internal frame accumulator.
Forwarding the frame accumulator phase to specific tones requires setting metadata inputs in  \li{PulseData}.
The structure of the metadata inputs is discussed in~\ref{sec:metadata} and outlined in Table~\ref{tab:bitmasks}.
In Example~\ref{ex:framefwdinv}, frame 0 is incremented by 15 degrees in each call to \li{PulseData}, but the the metadata bits in \li{fwd_frame0_mask} and \li{inv_frame0_mask} are used to control how the accumulated phase in frame 0 can be inverted and forwarded to the two tones.
\begin{example}[H]
\vspace*{1em}

  \begin{minipage}[t]{.58\textwidth}
  \vspace{0pt}
  \code{code/pulsedata/gpFrameRotFwdAndInv.py.tex}
  \end{minipage}%
  \begin{minipage}[t]{0.3\textwidth}
  \vspace{0pt}\raggedright

    \begin{tikzpicture}
            \pgfplotsset{ 
                every axis legend/.append style={at={(0,1)},anchor=north west,font=\scriptsize},
                scale only axis,
                xmin=-0.5,xmax=3.0,
                ymin=-49.5,ymax=49.50000000000001,
                every axis plot/.append style={line join=round,line cap=round,clip=false}
            }

            \begin{axis}[
                width=\linewidth,
                xlabel=Time ($\mu$s),
                ylabel=Applied Frame 0 (deg.),
                samples=100,
                yticklabels={-45,-30,-15,0,15,30,45},
                ytick={-45.0,-30.0,-15.0,0.0,15.0,30.0,45.0},
                xticklabels={0.00,1.00,2.00,3.00},
                xtick={0.0,1.0,2.0,3.0}
                ]
    \addplot[const plot, draw=black] coordinates {(0.0,0.0) (0.0,15.0) (1.0,0.0)};
\addplot[const plot, draw=black, forget plot] coordinates {(1.0,0.0) (2.0,-45.0)};
\addplot[const plot, draw=black, forget plot] coordinates {(2.0,-45.0) (3.0,-45.0)};
\addlegendentry{tone 0}
\addplot[const plot, draw=black,dashed] coordinates {(0.0,0.0) (1.0,-30.0)};
\addplot[const plot, draw=black,dashed, forget plot] coordinates {(1.0,-30.0) (2.0,45.0)};
\addplot[const plot, draw=black,dashed, forget plot] coordinates {(2.0,45.0) (3.0,45.0)};
\addlegendentry{tone 1}
\addplot[const plot, draw=black,dotted] coordinates {(-1.0,0.0) (0.0, 15.0) (1.0,30.0) (2.0,45.0) (3.0,45.0)};
\addlegendentry{frame 0}
        \end{axis}
    \end{tikzpicture}
    
  \end{minipage}
  \hfill
\caption{\label{ex:framefwdinv}
Frames can be optionally applied by forwarding them to one or both tones.
Likewise, their sign can be inverted on a per-tone basis.
}
\end{example}
Note that both frame 0 and frame 1 can be forwarded during the same pulse.
However, frame 0 will take precedence if both frames are forwarded to the same tone.

There are a couple more metadata options for frame rotations.
Mask settings for these bits are frame specific (\li{frame0} and \li{frame1}) and not tone specific (\li{tone0} and \li{tone1}).
The first option is \li{apply_at_end_mask}, which will update the frame for the {\it next} pulse.
This is useful in situations such as accounting for AC Stark effects, where adding a small phase shift in the rotating frame of the qubit after the pulse ends is necessary for global phase synchronization.
The second option is \li{rst_frame_mask}, which will clear the accumulated phase in the frame.
\li{rst_frame_mask} always happens at the beginning of a pulse, so a simultaneous application of \li{rst_frame_mask} and \li{apply_at_end_mask} will clear the frame for the current pulse, and the frame will take on the \li{frame} input value with the next pulse.

\begin{example}[H]
\vspace*{1em}

  \begin{minipage}[t]{.58\textwidth}
  \vspace{0pt}
  \code{code/pulsedata/gpFrameRotApplyAtEnd.py.tex}
  \end{minipage}%
  \begin{minipage}[t]{0.3\textwidth}
  \vspace{0pt}\raggedright

    \begin{tikzpicture}
            \pgfplotsset{ 
                scale only axis,
                xmin=-0.5,xmax=3.5,
                ymin=-6.0,ymax=11.0,
                every axis plot/.append style={line join=round,line cap=round,clip=false}
            }

            \begin{axis}[
                width=\linewidth,
                xlabel=Time ($\mu$s),
                ylabel=Frame 0 (deg.),
                samples=100,
                xticklabels={0.00,1.00,2.00,3.00},
                xtick={0.0,1.0,2.0,3.0}
                ]
    \addplot[const plot, draw=black] coordinates {(0.0,0.0) (1.0,10.0)};
\addplot[const plot, draw=black, forget plot] coordinates {(1.0,10.0) (2.0,-5.0)};
\addplot[const plot, draw=black, forget plot] coordinates {(2.0,-5.0) (3.0,-5.0)};
\addplot[const plot, draw=black,dashed, forget plot] coordinates {(3.0,-5.0) (4.0,-5.0)};
        \end{axis}
    \end{tikzpicture}
    
  \end{minipage}
  \hfill
\caption{\label{ex:frameapplyatend}
Frame rotations can optionally be applied with the {\it next} pulse, and the frames can also be reset.
}
\end{example}

Frame rotations also accept spline inputs.
In this case, the \li{apply_at_end_mask} is ignored, and the spline will effectively start from the initial value of the accumulator and the final value of the spline will be added to the frame accumulator.

\begin{example}[H]
\vspace*{1em}

  \begin{minipage}[t]{.58\textwidth}
  \vspace{0pt}
  \code{code/pulsedata/gpFrameRotSpline.py.tex}
  \end{minipage}%
  \begin{minipage}[t]{0.3\textwidth}
  \vspace{0pt}\raggedright

    \begin{tikzpicture}
            \pgfplotsset{ 
                scale only axis,
                xmin=-0.5,xmax=5.5,
                ymin=-0.5,ymax=27.500000000000004,
                every axis plot/.append style={line join=round,line cap=round,clip=false}
            }

            \begin{axis}[
                width=\linewidth,
                xlabel=Time ($\mu$s),
                ylabel=Frame 0 (deg.),
                samples=100,
                xticklabels={0.00,1.00,2.00,3.00,4.00,5.00},
                xtick={0.0,1.0,2.0,3.0,4.0,5.0}
                ]
    \addplot[const plot, draw=black] coordinates {(0.0,15.0) (1.0,15.0)};
\addplot[draw=black, forget plot][domain = 1.0:2.0]{-9.66667*((x-1.0)/(1.0))^3+0.00000*((x-1.0)/(1.0))^2+19.66667*((x-1.0)/(1.0))^1+15.00000*((x-1.0)/(1.0))^0};
\addplot[draw=black, forget plot][domain = 2.0:3.0]{18.33333*((x-2.0)/(1.0))^3+-29.00000*((x-2.0)/(1.0))^2+-9.33333*((x-2.0)/(1.0))^1+25.00000*((x-2.0)/(1.0))^0};
\addplot[draw=black, forget plot][domain = 3.0:4.0]{-8.66667*((x-3.0)/(1.0))^3+26.00000*((x-3.0)/(1.0))^2+-12.33333*((x-3.0)/(1.0))^1+5.00000*((x-3.0)/(1.0))^0};
\addplot[const plot, draw=black, forget plot] coordinates {(4.0,10.0) (5.0,10.0)};
        \end{axis}
    \end{tikzpicture}
    
  \end{minipage}
  \hfill
\caption{\label{ex:framespline}
Frame rotations support spline inputs.
Only the final value of the spline is added to the accumulator.
}
\end{example}

While this might not seem immediately useful, accounting for AC Stark shifts via splines offers certain advantages.
For example, the frequency offset associated with an AC Stark shift as a function of pulse amplitude can be represented as the integral of the pulse shape and normalized to the total phase offset from the resulting Stark shift.
The benefits in this case are twofold: 
\begin{itemize}
\item Time-dependent frequency variations are automatically accounted for via the time-dependent phase, allowing one to more closely match resonance conditions during the entire pulse.
\item Using frequencies in the natural (unshifted) frame of the qubit greatly simplifies the use of the global synchronization capabilities of the Octet.
\end{itemize}

\subsection{Metadata\label{sec:metadata}}

Each \pd object contains a number of entries designed to control certain operations or state associated with the pulse.
The list of metadata parameters is shown in Table~\ref{tab:metadatainputs}.
All metadata parameters are listed for the sake of completeness, but users should beware of using certain inputs.
Parameters that should generally be avoided are \li{enable_mask} and \li{waittrig}.
\li{enable_mask} will change the default output state to off, resulting in a zero output.
\li{waittrig} is used for handshaking with other control hardware to ensure that sequences run after other experimental processes have reached the correct state.

\begin{table}[h]
  \begin{tabular}{|c|c|c|c|}
    \hline
    Parameter              & Type & Default & Description \\
    \hline
    \hline
    \text{\li{sync_mask}}         & int  & 0b00    & Synchronize phase for current frequency \\
    \hline
    \text{\li{enable_mask}}       & int  & 0b00    & Toggle the output enable state \\
    \hline
    \text{\li{fb_enable_mask}}    & int  & 0b00    & Enable frequency correction \\
    \hline
    \text{\li{fwd_frame0_mask}}   & int  & 0b00    & Forward frame 0 \\
    \hline
    \text{\li{fwd_frame1_mask}}   & int  & 0b00    & Forward frame 1 \\
    \hline
    \text{\li{inv_frame0_mask}}   & int  & 0b00    & Invert frame 0 sign \\
    \hline
    \text{\li{inv_frame1_mask}}   & int  & 0b00    & Invert frame 1 sign  \\
    \hline
    \hline
    \text{\li{apply_at_end_mask}} & int  & 0b00    & Apply frame rotation at end of pulse \\
    \hline
    \text{\li{rst_frame_mask}}    & int  & 0b00    & Reset accumulated frame rotation \\
    \hline
    \hline
    \text{\li{waittrig}}          & bool & False   & Wait for external trigger \\
    \hline
  \end{tabular}
  \caption{\label{tab:metadatainputs}
  Metadata inputs for \pd objects. 
  All inputs ending in \li{_mask} range from 0-3 and follow a bitwise convention shown in table~\ref{tab:bitmasks}.
  The format \li{0b11} is Python notation for a base 2 integer, which in this case corresponds to a decimal value of 3.
  }
\end{table}

Metadata inputs ending with \li{_mask} take in tone-specific boolean values in the form of an integer.
This integer input is meant to be interpreted as a binary bit mask that is \li{True} if the bit is 1, and \li{False} if the bit is 0.
The least significant bit (LSB) corresponds to tone 0, and the most significant bit (MSB) corresponds to tone 1.

\begin{table}[H] 
  \centering
  \begin{tabular}{|c|c|c|}
    \hline
    Input & Tone 1 & Tone 0\\
    \hline 
    \hline
    0b00 & - & - \\
    \hline
    0b01 & - & \checkmark \\
    \hline
    0b10 & \checkmark & - \\
    \hline
    0b11 & \checkmark & \checkmark \\
    \hline
  \end{tabular}
  \caption{\label{tab:bitmasks}
  Convention used for bitmask enables. The least significant bit (LSB) controls tone 0, the most significant bit (MSB) controls tone 1.
  The same convention is used for \li{rst_frame_mask} and \li{apply_at_end_mask}, however the bits apply to frame 0 and frame 1 for the LSB and MSB respectively. 
  }
\end{table}

Most of the metadata is specific to frame rotations, covered in section~\ref{sec:framerotations}.
This leaves two of the most critical metadata entries, \li{sync_mask} and \li{fb_enable_mask}.
Improper application of these inputs, or lack thereof, can lead to unexpected behavior.
As such, these two features have been given dedicated sections.

\section{Frequency Feedback\label{sec:freqfb}}

The frequency stabilization techniques used are detailed in the QSCOUT manual.
A quick summary of this aspect of the system will be provided for completeness.
The Raman transitions used for gates in QSCOUT are driven by a pulsed laser.
This laser generates a frequency comb in which the comb teeth are separated by $\approx$ 120 MHz.
In order to drive the qubit transition, one can make use of a harmonic of the comb that approximately bridges the $\approx$12.642 GHz transition frequency.
For 120 MHz, this is roughly the 105$^{\text{th}}$ harmonic, where $105\times120~\text{MHz} = 12.6~\text{GHz}$.
This leaves a $\approx$ 42 MHz offset from resonance, which can easily be accounted for with an acousto-optic modulator (AOM).

The frequencies of the two tones applied for a gate must bridge this leftover 42 MHz frequency difference\footnote{In the actual QSCOUT platform, the numbers are slightly different and the resulting offset is closer to 28 MHz, but should be referenced from calibration data.}.
However, the pulsed laser used does not have active stabilization of the cavity length.
Thermal variation will cause the cavity length to drift, affecting the pulse repetition rate and thus the spacing of the frequency comb.
This ``breathing'' of the frequency comb requires that we actively correct for the resulting frequency offset between the two comb teeth needed to achieve resonance.
The variation in this repetition rate is constantly tracked, and we translate this variation into the correct frequency offset which is then forwarded to an output tone.

This frequency feedforward scheme is robust against the level of drift we normally encounter, and we've been able to yield coherence times exceeding 10 s~\cite{QSCOUTManual}.
However, it only works if we offset a {\it single} tone in the Raman transition.
In principle, we could add equal and opposite corrections to each tone, with half the total correction being applied.
But the sign must differ between the two cases and poses additional challenges based on the configuration of the tones.
This correction needs to be properly scaled to account for the total frequency variation at the particular harmonic to which we lock.
While we can apply the correction to either tone, this results in a sign change to the scaling factor.

The settings for the harmonic scaling and their sign are currently not reconfigurable between gates.
Rather, we've chosen a particular sign convention where the tone which is higher in frequency\textemdash in the case of the example numbers given above, this would be 42 MHz higher\textemdash should have \li{fb_enable_mask} set high.
The following code demonstrates how co-propagating and counter-propagating gates might implement the feedback enable.

\vspace*{2em}
\begin{lstlisting}
def gate_G_coprop(self, qubit):
    return [PulseData(qubit,       1e-6, freq0=200e6, 
                                         freq1=242e6, fb_enable_mask=0b10)]

def gate_G_counterprop(self, qubit):
    return [PulseData(qubit,       1e-6, freq0=200e6, fb_enable_mask=0b00),
            PulseData(GLOBAL_BEAM, 1e-6, freq0=242e6, fb_enable_mask=0b01)]
\end{lstlisting}

\section{Global Phase Synchronization}

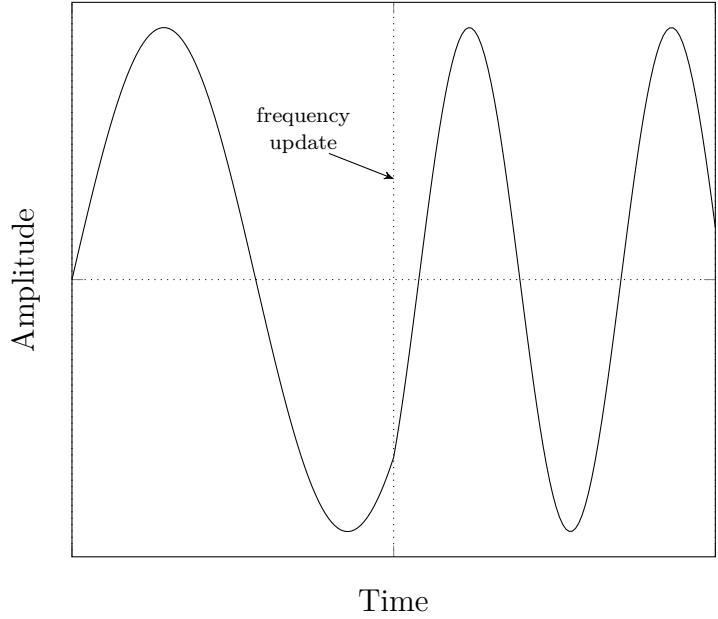
\begin{figure}
  \begin{minipage}[t]{.65\textwidth}
  \vspace{0pt}

\begin{tikzpicture}
		\pgfplotsset{
			scale only axis,
			major grid style={dotted,black},
			xmin=0.0,xmax=2.0,
			ymin=-1.1,ymax=1.1,
            every axis plot/.append style={line join=round,line cap=round,clip=false}
		}

		\begin{axis}[
                width=0.8\linewidth,
				tick style={grid=major},
				xlabel=Time,
				ylabel=Amplitude,
				samples=100,
				xticklabels={,,},
				xtick={0,1,2},
				ytick={-2,0,2},
				yticklabels={,,}
			]
\addplot[][domain = 0:1]{sin(5.5*deg(x))};
\addplot[][domain = 1:2]{sin(10.0*deg(x-1)+5.5*deg(1.0))};

\addplot[draw=black,->,>=stealth', forget plot] coordinates {(0.8,0.5) (1.0,0.4)}
        [pos segment=0,yshift=10pt,xshift=-0pt,font=\scriptsize] node[pos=-.4] {frequency}
        [pos segment=0,yshift=-10pt,xshift=-0pt,font=\scriptsize] node[pos=-.4] {update};
	\end{axis}
\end{tikzpicture}
  \end{minipage}%
  \caption{\label{fig:contFreqUpdate}
  Waveform output after a change in frequency.
  The phase is continuous at the boundary where the update occurs.
  }
\end{figure}

One of the most critical aspects of implementing a proper gate is ensuring that the phase of the gate is set correctly with respect to the qubit(s).
In many cases, the frequencies associated with a gate are subject to change.
For example, a two-qubit gate needs to address motional sidebands of the qubit transition, and certain gates may implement frequency modulation.
However, each hardware channel is equipped with only two custom direct digital synthesizer (DDS) modules for the two output tones.
Phase is continuous at the boundary where a frequency update occurs, as shown in Fig.~\ref{fig:contFreqUpdate}.
The implication is that the phase is generally arbitrary for the new frequency.

\begin{figure}
  \begin{minipage}[t]{.65\textwidth}
  \vspace{0pt}

\begin{tikzpicture}
		\pgfplotsset{
			scale only axis,
			major grid style={dotted,black},
			xmin=0.0,xmax=3.0,
			ymin=-1.1,ymax=1.1,
            every axis plot/.append style={line join=round,line cap=round,clip=false}
		}

		\begin{axis}[
            width=0.8\linewidth,
			tick style={grid=major},
			ylabel=Amplitude,
			samples=100,
            xticklabels={$f=f_0$,$f\rightarrow f_1$,$f\rightarrow f_0$,},
            xtick={0,1,2,3},
			axis x line*=top,
            ytick={-2,0,2},
            yticklabels={,}
			]
\addplot[draw=black,dashed][domain = 1:3]{sin(5.5*deg(x))};
\addplot[][domain = 0:1]{sin(5.5*deg(x))};
\addplot[][domain = 1:2]{sin(10.0*deg(x-1)+5.5*deg(1.0))};
\addplot[][domain = 2:3]{sin(5.5*deg(x-2)+10.0*deg(1.0)+5.5*deg(1.0))};

		\end{axis}
		\begin{axis}[
            width=0.8\linewidth,
			xlabel=Time,
			xmin=0.0,xmax=3.0,
			ymin=-1.1,ymax=1.1,
			xtick={0,1,2,3},
            xticklabels={,,,},
            ytick={-2,0,2},
            yticklabels={,,}
			]
	\end{axis}
\end{tikzpicture}
  \end{minipage}
  \caption{\label{fig:contFreqUpdateReturn}
  Waveform output after changing frequency and subsequently returning to the original frequency at a later time.
  The final phase is not aligned to the that of the original waveform.
  }
\end{figure}
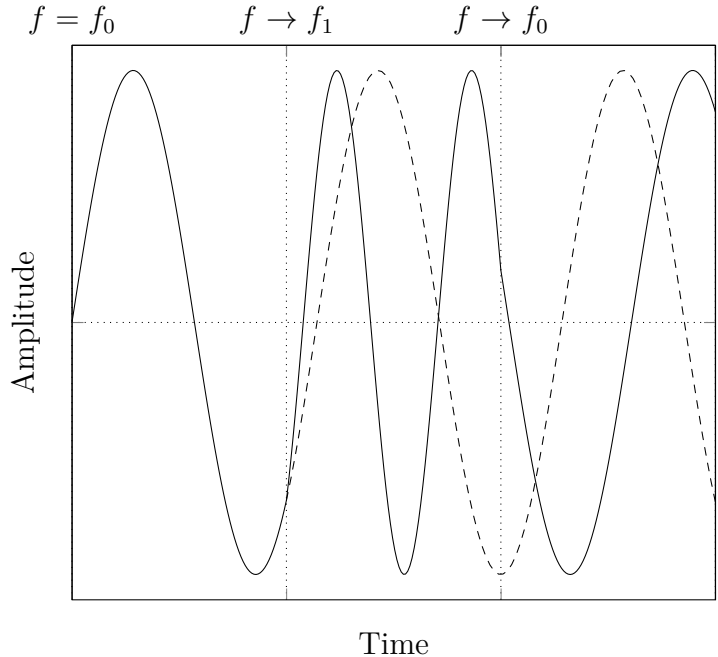

The absolute phase of the rf output is irrelevant from the perspective of a qubit initialized to an energy eigenstate.
However, the first gate applied to an initialized qubit sets the phase from which all subsequent gates need to be referenced during a circuit.
Thus a free-running synthesizer at a fixed frequency on resonance with the qubit transition can easily perform single-qubit gates about different axes by adjusting their phase relative to some initial value.
If the frequency is temporarily changed in order to perform another operation, then returning to the original frequency will typically result in a phase offset from the original waveform as shown in Fig.~\ref{fig:contFreqUpdateReturn}.

The resulting phase depends on the value of the other frequency and the duration for which it was applied.
While the phase offset can be easily calculated to determine a correction for returning to the original phase at the original frequency, this requires a lot of phase bookkeeping and is strongly context dependent.
The Octet hardware handles phase bookkeeping automatically via a modification to the standard DDS design.

\begin{figure}
  \begin{minipage}[t]{.65\textwidth}
  \vspace{0pt}

\begin{tikzpicture}
		\pgfplotsset{
			scale only axis,
                        major grid style={dotted,black},
			xmin=0.0,xmax=3.0,
			ymin=-1.1,ymax=1.1,
            every axis plot/.append style={line join=round,line cap=round,clip=false}
		}

		\begin{axis}[
                        width=0.8\linewidth,
			tick style={grid=major},
			ylabel=Amplitude,
			samples=100,
                        xticklabels={$f=f_0$,$f\rightarrow f_1$,\qquad~$f\rightarrow f_0\text{, sync}$,},
                        xtick={0,1,2,3},
			axis x line*=top,
                        ytick={-2,0,2},
                        yticklabels={,,}
			]
        \addplot[draw=black,dashed][domain = 1:2]{sin(5.5*deg(x))};
        \addplot[][domain = 0:1]{sin(5.5*deg(x))};
        \addplot[][domain = 1:2]{sin(10.0*deg(x-1)+5.5*deg(1.0))};
        \addplot[][domain = 2:3]{sin(5.5*deg(x)};
        \addplot[draw=black] coordinates {(2.0,{sin(5.5*deg(2.0)}) (2.0,{sin(10.0*deg(1)+5.5*deg(1.0))})};
        \end{axis}
        \begin{axis}[
                width=0.8\linewidth,
	        xlabel=Time,
	        xmin=0.0,xmax=3.0,
		ymin=-1.1,ymax=1.1,
		xtick={0,1,2,3},
                xticklabels={,,,},
                ytick={-2,0,2},
                yticklabels={,,}
		]
	\end{axis}
\end{tikzpicture}
  \end{minipage}
  \caption{\label{fig:freqUpdateWithSync}
  Waveform output after a change in frequency, followed by a return to the original frequency with a synchronization call applied.
  In this case, the waveform experiences a discontinuous phase jump when returning to the original frequency {\it and} phase.
  }
\end{figure}
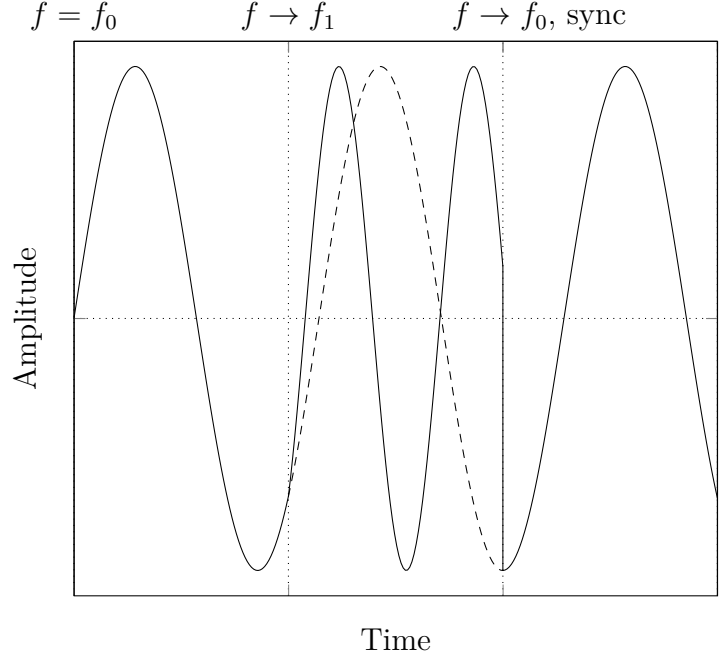

A global counter shared by all output tones is constantly being multiplied by the DDS modules input frequency word.
This produces a global phase, $\Phi = \omega t$, where $\omega$ is the input frequency, and the time, $t$, is the value of the the global counter.
The frequency input to the DDS is delayed to match the latency of the multiplication stage such that the global phase tracks with the input frequency.
Setting \li{sync_mask} high for a particular tone in a \pd object will send a single trigger at the beginning of the pulse.
This trigger will overwrite the DDS phase accumulator with the global phase, $\Phi$.
The resulting behavior is equivalent to the second frequency update in Fig.~\ref{fig:freqUpdateWithSync}. 

It is worth noting that this mechanism functions the same, regardless of the other settings.
In other words, synchronization is not affected by other metadata such as \li{fb_enable_mask} or settings on the other tone.
Synchronization only depends on the frequency at the start of the pulse\footnote{For frequency modulation, the synchronization step is applied with the beginning of the first element in the modulation list or tuple.}.

\subsection{Synchronization Caveats}

Gates that require higher-order frequency and phase relationships might require some extra consideration.
For example, the M\o lmer-S\o rensen gate requires that the red and blue sideband frequencies obey the relationship
\begin{equation}
   f_r + f_b = 2f_{qubit}
\end{equation}
where
\begin{align}
  f_r &\equiv f_{qubit} - f_{\textsc{\tiny SB}} \\
  f_b &\equiv f_{qubit} + f_{\textsc{\tiny SB}}. \\
\end{align}
This condition maps equivalently onto phase such that
\begin{equation}
  \phi_r + \phi_b = 2\phi_{qubit}.
\end{equation}
Because frequencies are digitized before being sent to hardware, rounding effects can result in a small difference in the final frequencies such that
\begin{equation}
  F_r + F_b \neq 2F_{qubit},
\end{equation}
where $F$ indicates the frequency word in the digital domain.
This phase relationship is difficult to predict from the vantage point of the compiler.
It is up to the user to ensure that the desired frequency relationships match in the digital domain to ensure proper phase relationships between tones.

These rounding errors typically contribute a 0 or 1 bit difference from the desired result, potentially adding a frequency offset of $\approx 745 \mu\text{Hz}$.
While this may seem negligible, it's important to understand that the global counter is generally set to zero when the hardware is powered on.
In some cases, the counter may be accumulating for weeks or months.
The global phase error between $F_{qubit}$ and $F_{qubit}+F_{eps}$ is given by $\delta\phi = F_{eps}t_{global}$, which can be quite substantial when $t_{global} >> 0$.


Helper functions are provided to simplify the calculations.
For synchronization purposes, \li{discretize_frequency} can be imported from \li{jaqalpaw.utilities.helper_functions}.
Similar functions \li{discretize_amplitude} and \li{discretize_phase} are also provided and can be imported from the same path.
The usage for \li{discretize_frequency} is shown in Ex.~\ref{ex:discretizefreq} and involves converting frequencies to the digital domain before calculating their frequency relationships.

In this case, the resonant qubit frequency is discretized, as well as the desired sideband frequency, prior to the red and blue sideband frequency calculations.
The frequency inputs in the \pd call take \li{rsb_freq} and \li{bsb_freq}, which have been calculated using discretized frequencies.
Note that the discretization effects need only be taken into account for the relevant leg of the Raman transition, which in this case is given by \li{self.aom_center_frequency + self.effective_qubit_splitting}.

\begin{example}[H]
\begin{minipage}[b]{1.0\textwidth}
\codeopts{xleftmargin=0.05\textwidth,
          xrightmargin=.05\textwidth, 
          frame=single}{code/synchronization/discretize_frequency_example.py}
\end{minipage}
\caption{\label{ex:discretizefreq}
A basic gate pulse class.
}
\end{example}

\bibliography{IonTrapBibliography}

\begin{thebibliography}{4}%
\makeatletter
\providecommand \@ifxundefined [1]{%
 \@ifx{#1\undefined}
}%
\providecommand \@ifnum [1]{%
 \ifnum #1\expandafter \@firstoftwo
 \else \expandafter \@secondoftwo
 \fi
}%
\providecommand \@ifx [1]{%
 \ifx #1\expandafter \@firstoftwo
 \else \expandafter \@secondoftwo
 \fi
}%
\providecommand \natexlab [1]{#1}%
\providecommand \enquote  [1]{``#1''}%
\providecommand \bibnamefont  [1]{#1}%
\providecommand \bibfnamefont [1]{#1}%
\providecommand \citenamefont [1]{#1}%
\providecommand \href@noop [0]{\@secondoftwo}%
\providecommand \href [0]{\begingroup \@sanitize@url \@href}%
\providecommand \@href[1]{\@@startlink{#1}\@@href}%
\providecommand \@@href[1]{\endgroup#1\@@endlink}%
\providecommand \@sanitize@url [0]{\catcode `\\12\catcode `\$12\catcode
  `\&12\catcode `\#12\catcode `\^12\catcode `\_12\catcode `\%12\relax}%
\providecommand \@@startlink[1]{}%
\providecommand \@@endlink[0]{}%
\providecommand \url  [0]{\begingroup\@sanitize@url \@url }%
\providecommand \@url [1]{\endgroup\@href {#1}{\urlprefix }}%
\providecommand \urlprefix  [0]{URL }%
\providecommand \Eprint [0]{\href }%
\providecommand \doibase [0]{https://doi.org/}%
\providecommand \selectlanguage [0]{\@gobble}%
\providecommand \bibinfo  [0]{\@secondoftwo}%
\providecommand \bibfield  [0]{\@secondoftwo}%
\providecommand \translation [1]{[#1]}%
\providecommand \BibitemOpen [0]{}%
\providecommand \bibitemStop [0]{}%
\providecommand \bibitemNoStop [0]{.\EOS\space}%
\providecommand \EOS [0]{\spacefactor3000\relax}%
\providecommand \BibitemShut  [1]{\csname bibitem#1\endcsname}%
\let\auto@bib@innerbib\@empty
\bibitem [{\citenamefont {Clark}\ \emph {et~al.}(2021)\citenamefont {Clark},
  \citenamefont {Lobser}, \citenamefont {Revelle}, \citenamefont {Yale},
  \citenamefont {Bossert}, \citenamefont {Burch}, \citenamefont {Chow},
  \citenamefont {Hogle}, \citenamefont {Ivory}, \citenamefont {Pehr},
  \citenamefont {Salzbrenner}, \citenamefont {Stick}, \citenamefont {Sweatt},
  \citenamefont {Wilson}, \citenamefont {Winrow},\ and\ \citenamefont
  {Maunz}}]{QSCOUTManual}%
  \BibitemOpen
  \bibfield  {author} {\bibinfo {author} {\bibfnamefont {S.~M.}\ \bibnamefont
  {Clark}}, \bibinfo {author} {\bibfnamefont {D.}~\bibnamefont {Lobser}},
  \bibinfo {author} {\bibfnamefont {M.}~\bibnamefont {Revelle}}, \bibinfo
  {author} {\bibfnamefont {C.~G.}\ \bibnamefont {Yale}}, \bibinfo {author}
  {\bibfnamefont {D.}~\bibnamefont {Bossert}}, \bibinfo {author} {\bibfnamefont
  {A.~D.}\ \bibnamefont {Burch}}, \bibinfo {author} {\bibfnamefont {M.~N.}\
  \bibnamefont {Chow}}, \bibinfo {author} {\bibfnamefont {C.~W.}\ \bibnamefont
  {Hogle}}, \bibinfo {author} {\bibfnamefont {M.}~\bibnamefont {Ivory}},
  \bibinfo {author} {\bibfnamefont {J.}~\bibnamefont {Pehr}}, \bibinfo {author}
  {\bibfnamefont {B.}~\bibnamefont {Salzbrenner}}, \bibinfo {author}
  {\bibfnamefont {D.}~\bibnamefont {Stick}}, \bibinfo {author} {\bibfnamefont
  {W.}~\bibnamefont {Sweatt}}, \bibinfo {author} {\bibfnamefont {J.~M.}\
  \bibnamefont {Wilson}}, \bibinfo {author} {\bibfnamefont {E.}~\bibnamefont
  {Winrow}},\ and\ \bibinfo {author} {\bibfnamefont {P.}~\bibnamefont
  {Maunz}},\ }\href@noop {} {\bibinfo {title} {Engineering the quantum
  scientific computing open user testbed (qscout): Design details and user
  guide}} (\bibinfo {year} {2021}),\ \Eprint {https://arxiv.org/abs/in
  preparation} {in preparation} \BibitemShut {NoStop}%
\bibitem [{\citenamefont {Landahl}\ \emph {et~al.}(2020)\citenamefont
  {Landahl}, \citenamefont {Lobser}, \citenamefont {Morrison}, \citenamefont
  {Rudinger}, \citenamefont {Russo}, \citenamefont {{Van Der Wall}},\ and\
  \citenamefont {Maunz}}]{Landahl2020}%
  \BibitemOpen
  \bibfield  {author} {\bibinfo {author} {\bibfnamefont {A.~J.}\ \bibnamefont
  {Landahl}}, \bibinfo {author} {\bibfnamefont {D.~S.}\ \bibnamefont {Lobser}},
  \bibinfo {author} {\bibfnamefont {B.~C.~A.}\ \bibnamefont {Morrison}},
  \bibinfo {author} {\bibfnamefont {K.~M.}\ \bibnamefont {Rudinger}}, \bibinfo
  {author} {\bibfnamefont {A.~E.}\ \bibnamefont {Russo}}, \bibinfo {author}
  {\bibfnamefont {J.~W.}\ \bibnamefont {{Van Der Wall}}},\ and\ \bibinfo
  {author} {\bibfnamefont {P.}~\bibnamefont {Maunz}},\ }\href@noop {} {\bibinfo
  {title} {Jaqal, the quantum assembly language for qscout}} (\bibinfo {year}
  {2020}),\ \Eprint {https://arxiv.org/abs/2003.09382} {arXiv:2003.09382
  [quant-ph]} \BibitemShut {NoStop}%
\bibitem [{\citenamefont {Morrison}\ \emph {et~al.}(2020)\citenamefont
  {Morrison}, \citenamefont {Landahl}, \citenamefont {Lobser}, \citenamefont
  {Rudinger}, \citenamefont {Russo}, \citenamefont {{Van Der Wall}},\ and\
  \citenamefont {Maunz}}]{Morrison2020}%
  \BibitemOpen
  \bibfield  {author} {\bibinfo {author} {\bibfnamefont {B.~C.~A.}\
  \bibnamefont {Morrison}}, \bibinfo {author} {\bibfnamefont {A.~J.}\
  \bibnamefont {Landahl}}, \bibinfo {author} {\bibfnamefont {D.~S.}\
  \bibnamefont {Lobser}}, \bibinfo {author} {\bibfnamefont {K.~M.}\
  \bibnamefont {Rudinger}}, \bibinfo {author} {\bibfnamefont {A.~E.}\
  \bibnamefont {Russo}}, \bibinfo {author} {\bibfnamefont {J.~W.}\ \bibnamefont
  {{Van Der Wall}}},\ and\ \bibinfo {author} {\bibfnamefont {P.}~\bibnamefont
  {Maunz}},\ }\href@noop {} {\bibinfo {title} {Just another quantum assembly
  language (jaqal)}} (\bibinfo {year} {2020}),\ \Eprint
  {https://arxiv.org/abs/2008.08042} {arXiv:2008.08042 [quant-ph]} \BibitemShut
  {NoStop}%
\bibitem [{Pyt()}]{Python3Dataclass}%
  \BibitemOpen
  \href@noop {} {\bibinfo {title} {Python data classes}},\ \bibinfo
  {howpublished} {\url{https://docs.python.org/3/library/dataclasses.html}},\
  \bibinfo {note} {[Online; accessed 6-April-2021]}\BibitemShut {NoStop}%
\end{thebibliography}%
\end{document}